\documentclass[aps,prd,preprint,superscriptaddress,showpacs]{revtex4}
\usepackage{amsmath}
\usepackage{amssymb}
\usepackage{epsfig}
\usepackage{ulem}
\usepackage{graphics}
\usepackage{indentfirst}
\usepackage{color}
\usepackage{multirow}

\newcommand{\tb}{\ensuremath{\bar{T}}}

\newcommand{\pb}{\ensuremath{\bar{\Phi}}}

\newcommand{\te}{\ensuremath{\theta}}
\newcounter{RomanNumber}
\newcommand{\MyRoman}[1]{\setcounter{RomanNumber}{#1}\Roman{RomanNumber}}

\begin{document}

\preprint{ACT-13-14, MIFPA-14-39}

\vspace*{2cm}
\title{Helical Phase Inflation and Monodromy in Supergravity Theory \vspace*{1.5cm}}

\author{Tianjun Li \vspace*{1.2cm}}
\affiliation{{\footnotesize State Key Laboratory of Theoretical Physics
and Kavli Institute for Theoretical Physics China (KITPC),
      Institute of Theoretical Physics, Chinese Academy of Sciences,
Beijing 100190, P. R. China}}

\affiliation{{\footnotesize School of Physical Electronics,
University of Electronic Science and Technology of China,
Chengdu 610054, P. R. China}}

\author{Zhijin Li}

\affiliation{{\footnotesize George P. and Cynthia W. Mitchell Institute for
Fundamental Physics and Astronomy,
Texas A\&M University, College Station, TX 77843, USA}}

\author{Dimitri V. Nanopoulos}

\affiliation{{\footnotesize George P. and Cynthia W. Mitchell Institute for
Fundamental Physics and Astronomy,
Texas A\&M University, College Station, TX 77843, USA}}

\affiliation{{\footnotesize Astroparticle Physics Group, Houston Advanced
Research Center (HARC), Mitchell Campus, Woodlands, TX 77381, USA}}

\affiliation{{\footnotesize Academy of Athens, Division of Natural Sciences,
28 Panepistimiou Avenue, Athens 10679, Greece} \vspace*{0.3cm}}

\begin{abstract}
  We study helical phase inflation in supergravity theory in details. The inflation is driven by the phase component of a complex field along helical trajectory. The helicoid structure originates from the monodromy of superpotential with an singularity at origin.
  We show that such monodromy can be formed by integrating out heavy fields in supersymmetric field theory. The supergravity corrections to the potential provide strong field stabilizations for the scalars except inflaton, therefore the helical phase inflation accomplishes the ``monodromy inflation'' within supersymmetric field theory.
  The phase monodromy can be easily generalized for natural inflation, in which the super-Planckian phase decay constant is realized with consistent field stabilization. The phase-axion alignment is fulfilled indirectly in the process of integrating out the heavy fields.
  Besides, we show that the helical phase inflation can be naturally realized in no-scale supergravity with $SU(2,1)/SU(2)\times U(1)$ symmetry since the no-scale K\"ahler potential provides symmetry factors of phase monodromy directly. We also demonstrate that the helical phase inflation can reduce to the shift symmetry realization of supergravity inflation. The super-Planckian field excursion is accomplished by the phase component, which admits no dangerous polynomial higher order corrections. The helical phase inflation process is free from the UV-sensitivity problem, and it suggests that inflation can be effectively studied in supersymmetric field theory close to the unification scale in
Grand Unified Theory and a UV-completed frame is not prerequisite.
\end{abstract}


\maketitle

\section{Introduction}
Inflation theory plays a crucial role in the early stage of our Universe \cite{Guth:1980zm}.  Soon after the discovery of the inflation theory, supersymmetry was found to play an important role in it \cite{ENOT}. A general argument is the inflation process happens close to the unification scale in
Grand Unified Theory (GUT) \cite{Ade:2013uln, Ade:2014xna}, and at such scale physics theory is widely believed to be supersymmetric. Technically, to realize the slow-roll inflation, it requires the strict flat conditions for the potential $V(\phi)$ of inflaton $\phi$. The mass of inflaton $m_\phi$ should be significantly smaller than the inflation energy scale due to the slow-roll parameter
\begin{equation}
\eta\equiv M_P^2\frac{V''}{V}\simeq \frac{m_\phi^2}{3H^2}\ll1, \label{eta}
\end{equation}
where $M_P$ is the reduced Planck mass. Otherwise,  inflation cannot be triggered or last for a sufficient long period.
However, as a scalar field, the inflaton potential is expected to obtain large quantum loop corrections which break the slow-roll condition unless there is extremely fine tuning. Supersymmetry is a natural way to eliminate such quantum corrections, by introducing supersymmetry the flatness problem can be partially relaxed, but not completely solved since supersymmetry is broken during inflation.
Moreover, gravity plays an important role in inflation, so it is natural to study inflation within supergravity theory.

Once combing the supersymmetry and gravity theory together, the flatness problem reappears known as $\eta$ problem. $N=1$ supergravity in four-dimensional space-time is determined by three functions: K\"ahler potential $K$, superpotential $W$, and gauge kinetic function. The F-term scalar potential contains an exponential factor $e^K$. In the minimal supergravity with $K=\Phi\bar{\Phi}$, the exponential factor $e^K$ contributes a term in the inflaton mass at Hubble scale, which breaks the slow-roll condition (\ref{eta}). To realize inflation in supergravity, the large contribution to scalar mass from $e^K$ should be suppressed, which needs an extra symmetry. In the minimal supergravity, the $\eta$ problem can be solved by introducing shift symmetry in the K\"ahler potential as proposed by
Kawasaki, Yamaguchi, and Yanagida (KYY) \cite{Kawasaki:2000yn}: $K$ is invariant under the shift $\Phi\rightarrow \Phi+iC$. Consequently $K$ is independent of Im$(\Phi)$, so is the factor $e^K$ in the F-term potential. Employing Im$(\Phi)$ as inflaton, its mass is not affected by $e^K$ and then there is no $\eta$ problem any more. The shift symmetry can be slightly broken, in this case there is still no $\eta$ problem and the model gives a broad range of tensor-to-scalar ratio $\textbf{r}$ \cite{Li:2013nfa, Harigaya:2014qza}. The $\eta$ problem is automatically solved in no-scale supergravity because of the $SU(N,1)/SU(N)\times U(1)$ symmetry in the K\"ahler manifold. Historically, the no-scale supergravity was proposed to get vanishing cosmology constant \cite{Cremmer:1983bf}. At classical level the potential is strictly flat guaranteed by the $SU(N,1)/SU(N)\times U(1)$ symmetry of the K\"ahler potential, which meanwhile protects the no-scale type inflation away from $\eta$ problem. Moreover, the $SU(N,1)/SU(N)\times U(1)$ symmetry has rich structure that allows different types of inflation. Thus, the inflation based on no-scale supergravity has been extensively
 studied~\cite{Ellis:2013xoa,Li:2013moa, Pallis:2014dma, Ferrara:2014ima, Ellis:2014rxa, Li:2014owa, Ellis:2014gxa, Ferrara:2014fqa, Ketov:2014qha, Kounnas:2014gda, Diamandis:2014vxa, Pallis:2014cda}. In this work we will show that in no-scale supergravity with $SU(2,1)/SU(2)\times U(1)$ symmetry, one can pick up the $U(1)$ subsector, together with the superpotential phase monodromy to realize helical phase inflation.

Recently, it was shown that the $\eta$ problem can be naturally solved in helical phase inflation \cite{Li:2014vpa}.
This solution considers a global $U(1)$ symmetry, which is a trivial fact in the minimal supergravity with $K=\Phi\bar{\Phi}$. Using the phase of a complex field $\Phi$ as an inflaton, the $\eta$ problem is solved due to the global $U(1)$ symmetry. The norm of $\Phi$ needs to be stabilized otherwise it will generate notable iso-curvature perturbation that contradicts with observations. However, it is a non-trivial task to stabilize the norm of $\Phi$ while keep the phase light as the norm and phase couple with each other. In that work the field stabilization and quadratic
inflation are realized via a helicoid type potential. The inflationary trajectory is a helix line, and this is the reason for the name ``helical phase inflation''. In addition, the superpotential of helical phase inflation realizes monodromy in supersymmetric field theory. Furthermore, the helical phase inflation gives a method to avoid the dangerous quantum gravity effect on inflation.

The single field slow-roll inflation agrees with recent observations \cite{Ade:2013uln, Ade:2014xna}. Such kind of inflation admits a relationship between the inflaton field excursion and the tensor-to-scalar ratio, which is known as
the Lyth bound~\cite{Lyth:1996im}. It suggests that to get the large tensor-to-scalar ratio, the field excursion during inflation should be much larger than the Planck mass.  The super-Planckian field excursion challenges
the validity, in the Wilsonian sense, of inflationary models described by effective field theory. At the Planck scale the quantum gravity effect is likely to introduce extra terms which are suppressed by the Planck mass and then irrelevant in the low energy scale. While for a super-Planckian field, the irrelevant terms become important and may introduce significant corrections or even destroy the inflation process. In this sense  predictions just based on the effective field theory is not trustable. A more detailed discussion on the ultraviolet (UV) sensitivity of the inflation process is provided in the review \cite{Baumann:2014nda}.

A lot of works have been proposed to realize inflation based on the UV completed theory, for examples in
\cite{Marchesano:2014mla, Long:2014dta, Gao:2014uha, Ben-Dayan:2014lca, Blumenhagen:2014nba,
Abe:2014pwa, Ali:2014mra, Flauger:2014ana}. However, to realize inflation in string theory it needs to address several difficult problems such as moduli stabilization, Minkowski or de Sitter vacuum, $\alpha^\prime$- and higher string loop-corrections on the K\"ahler potential, etc. While one may doubt whether such difficult UV-completed framework
for inflation like  GUTs is necessary. In certain scenario the super-Planckian field excursion does not necessarily lead to the physical field above the Planck scale. A simple example is the phase of a complex field. The phase factor, like a pseudo-Nambu-Goldstone boson (PNGB), can be shifted to any value without any effect on the energy scale. By employing the phase as an inflaton, the super-Planckian field excursion is not problematic at all as there is no polynomial higher order quantum gravity correction for the phase component. Besides the helical phase inflation, inflationary models using PNGB as an inflaton have been studied~\cite{Freese:1990rb,Kim:2004rp,Baumann:2010nu,McDonald:2014oza, Barenboim:2014vea}. While for the natural inflation, it requires super-Planckian axion decay constant, which can be obtained by aligned axions \cite{Kim:2004rp} (the axion alignment relates to $S_n$ symmetry among K\"ahler moduli \cite{Li:2014lpa}) or anomalous $U(1)$ gauge symmetry with large condensation gauge group~\cite{Li:2014xna}. In helical phase inflation, as will be shown later, the phase monodromy in superpotential can be easily modified for natural inflation, and the super-Planckian phase decay constant can be realized, which is from the phase-axion alignment hidden in the process of integrating out heavy fields. Furthermore, all the extra fields are consistently stabilized based on the helicoid potential.

Like the helical phase inflation, the ``monodromy inflation'' was proposed to solve the UV sensitivity of large field inflation \cite{Silverstein:2008sg, Kaloper:2008fb}. In such model the inflaton is identified as an axion arising from $p-$form field after string compactifications.  The inflaton potential arises from the DBI action of branes or coupling between axion and fluxes. During inflation the axion transverses along internal cycles, for each cycle almost all the physical conditions stay the same except the potential.
An interesting realization of monodromy inflation is the axion alignment \cite{Kim:2004rp}, which was proposed to get super-Planckian axion decay constant for natural inflation, and it was noticed that this mechanism actually provides an axion monodromy in \cite{Choi:2014rja, Tye:2014tja, Kappl:2014lra}.
Actually a similar name ``helical inflation'' was firstly introduced in \cite{Tye:2014tja} for an inflation model with axion monodromy. However, a major difference should be noted, the ``helical'' structure in \cite{Tye:2014tja} is to describe the alignment structure of two axions, while the ``helical'' structure in our model is from a single complex field with stabilized field norm.
The physical picture of axion monodromy is analogical to the superpotential $W$ in helical phase inflation. For $W$ there is a monodromy, in the mathematical sense,  around the singularity $\Phi=0$,
\begin{equation}
\Phi\rightarrow \Phi e^{2\pi i}, ~~~~W\rightarrow W+2\pi i\frac{W}{\log\Phi}.
\end{equation}
The phase monodromy, together with the $U(1)$ symmetry in the K\"ahler potential, provides flat direction for inflation.
In the following we will show that this monodromy is corresponding to the global $U(1)$ symmetry explicitly broken by the inflation term.

In this work we will study the helical phase inflation from several aspects in details. Firstly, we will show that the helical structure of inflaton potential, which originates from the monodromy of superpotential at origin singularity, can be effectively generated by integrating out heavy fields in supersymmetric field theory.
Besides the quadratic inflation, the phase monodromy for helical phase inflation can be easily generalized to natural inflation, in which the process of integrating out heavy fields fulfills the phase-axion alignment indirectly and leads to super-Planckian phase decay constant with consistent field stabilization as well.
We also show that mathematically, the helical phase inflation can reduce to the KYY inflation by a field redefinition, however, there is no such field transformation that can map the KYY model back to the helical phase inflation.
Moreover, we show that the no-scale supergravity with $SU(2,1)/SU(2)\times U(1)$ symmetry provides a natural frame for helical phase inflation, as the $SU(2,1)/SU(2)\times U(1)$ symmetry of no-scale K\"ahler potential already combines the symmetry factors needed for phase monodromy. Moreover, we argue that the helical phase inflation is free from the UV-sensitivity problem.

This paper is organized as follows. In Section II we review the minimal supergravity construction of helical phase inflation. In Section III we present the realization of phase monodromy based on supersymmetric field theory.
In Section IV the natural inflation as a special type of helical phase inflation is studied.
In Section V the relationship between the helical phase inflation and the KYY model is discussed.
In Section VI we study the helical phase inflation in no-scale supergravity with $SU(2,1)/SU(2)\times U(1)$ symmetry.
In Section VII we discuss how the helical phase inflation dodges the UV-sensitivity problem of large field inflation. Conclusion is given in Section VIII.

\section{Helical phase inflation}

In four dimensions, $N=1$ supergravity is determined by the K\"ahler potential $K$, superpotential $W$ and gauge kinetic function. The F-term scalar potential is given by
\begin{equation}
V=e^K(K^{i\bar{j}}D_iW D_{\bar{j}}\bar{W}-3W\bar{W}).
\end{equation}
To realize inflation in supergravity, the factor $e^K$ in above formula is an obstacle as it makes the potential too steep for a sufficient long slow-roll process. This is the well-known $\eta$ problem.
To solve the $\eta$ problem usually one needs a symmetry in the K\"ahler potential. In the minimal supergravity,
  there is a global $U(1)$ symmetry in the K\"ahler $K=\Phi\bar{\Phi}$. This global $U(1)$ symmetry is employed in helical phase inflation. As the K\"ahler potential is independent on the phase $\theta$, the potential of phase $\theta$ is not affected by the exponential factor $e^K$, consequently, there is no $\eta$ problem for phase inflation. However, the field stabilization becomes more subtle. All the extra fields except inflaton have to be stabilized for single field inflation, but normally the phase and norm of a complex field couple with each other and then it is very difficult to stabilize norm while keep phase light.

The physical picture of helical phase inflation is that the phase evolves along a flat circular path with constant, or almost constant radius--the field magnitude, and the potential decreases slowly. So even before writing down the explicit supergravity formula, one can deduce that the phase inflation, if realizable, should be a particular realization of complex phase monodromy, and there exists an singularity in the superpotential that generates the phase monodromy.
Such singularity further indicates that the model is described by an effective theory.

The helical phase inflation is realized in the minimal supergravity with the K\"ahler potential
\begin{equation}
K=\Phi\pb+X\bar{X}-g(X\bar{X})^2, \label{kp}
\end{equation}
and superpotential
\begin{equation}
W=a\frac{X}{\Phi}\ln(\Phi). \label{sup}
\end{equation}
The global $U(1)$ symmetry in $K$ is broken by the superpotential with a small factor $a$, when $a\rightarrow0$ the $U(1)$ symmetry is restored. Therefore, the superpotential with small coefficient is technically natural \cite{tHooft}, which makes the model technically stable against radiative corrections.
As discussed before, the superpotential $W$ is singular at $\Phi=0$ and exhibits a phase monodromy
\begin{equation}
\Phi\rightarrow \Phi e^{2\pi i}, ~~W\rightarrow W+2\pi ai\frac{X}{\Phi}. \label{mono}
\end{equation}
The theory is well-defined only for $\Phi$ away from the singularity.

During inflation, the field $X$ is stabilized at $X=0$, and the scalar potential is simplified as
\begin{equation}
V=e^{\Phi\pb}W_X\bar{W}_{\bar{X}}=a^2e^{r^2}\frac{1}{r^2}((\ln r)^2+\te^2), \label{po}
\end{equation}
where $\Phi=re^{i\te}$, and the kinetic term is $L_K=\partial_\mu r\partial^\mu r+r^2\partial_\mu \te\partial^\mu \te$.
Interestingly, in the potential (\ref{po}), both the norm-dependent factor $e^{r^2}\frac{1}{r^2}$ and $(\ln r)^2$ reach the minimum at $r=1$. The physical mass of norm $r$ is
\begin{equation}
m_r^2=\frac{1}{2}\frac{\partial^2 V}{\partial r^2}|_{r=1}=(2+\frac{1}{\te^2})V_I,
\end{equation}
therefore the norm is strongly stabilized at $r=1$ during inflation and the Lagrangian for the inflaton is
\begin{equation}
L=\partial_\mu\te\partial^\mu \te-ea^2\te^2,
\end{equation}
which gives the quadratic inflation driven by the phase of complex field $\Phi$.

In the above simple example given by (\ref{kp}) and (\ref{sup}), the field stabilization is obtained from the combination of supergravity correction $e^K$ and the pole $\frac{1}{\Phi}$ in $W$, besides an accidental agreement that both the factor $\frac{1}{r^2}e^{r^2}$ and the term $(\ln r)^2$ obtain their minima at $r=1$. While for more general helical phase inflation, such accidental agreement is not guaranteed. For example, one may get the following
inflaton potential
\begin{equation}
V=e^{K(r)}\frac{1}{r^2}((\ln\frac{r}{\Lambda})^2+\te^2),
\end{equation}
in which the coefficient $e^{K(r)}\frac{1}{r^2}$ admits a minimum at $r_0\sim\Lambda$ but $r_0\neq\Lambda$. In this case, the coefficient $e^{K(r)}\frac{1}{r^2}$ still gives a mass above the Hubble scale for $r$, while $\langle r\rangle$ is slightly shifted away from $r_0$ in the early stage of inflation, and after inflation $r$ evolves to $\Lambda$ rapidly. Also, the term $(\ln \frac{r}{\Lambda})^2$ gives a small correction to the potential and inflationary observables,
so this correction is ignorable comparing with the contributions from the super-Planckian valued phase  unless it is unexpected large.

\subsection*{Potential Deformations}

In the K\"ahler potential there are corrections from the quantum loop effect, while the superpotential $W$ is non-renormalized. Besides, when coupled with heavy fields, the K\"ahler potential of $\Phi$ receives corrections through integrating out the heavy fields. Nevertheless, because of the global $U(1)$ symmetry in the K\"ahler potential, these corrections can only affect the field stabilization, while phase inflation is not sensitive to these corrections.

Giving a higher order correction on the K\"ahler potential
\begin{equation}
K=\Phi\pb+b(\Phi\pb)^2 + X\bar{X}-g(X\bar{X})^2,
\end{equation}
one may introduce an extra parameter $\Lambda$ in the superpotential
\begin{equation}
W=a\frac{X}{\Phi}\ln\frac{\Phi}{\Lambda}.
\end{equation}

Based on the same argument it is easy to see the scalar potential reduces to
\begin{equation}
V=a^2e^{r^2+br^4}\frac{1}{r^2}((\ln r-\ln\Lambda)^2+\te^2). \label{mini}
\end{equation}
The factor $e^{r^2+br^4}\frac{1}{r^2}$ reaches its minimum at $r^2_0=\frac{2}{1+{\sqrt {1+8b}}}$, below $M_P$ for $b>0$. To get the ``accidental agreement'' it needs the parameter $\Lambda=r_0$, and then inflation is still driven by the phase with exact quadratic potential.

\begin{figure}
\centering
\includegraphics[width=100mm, height=90mm,angle=0]{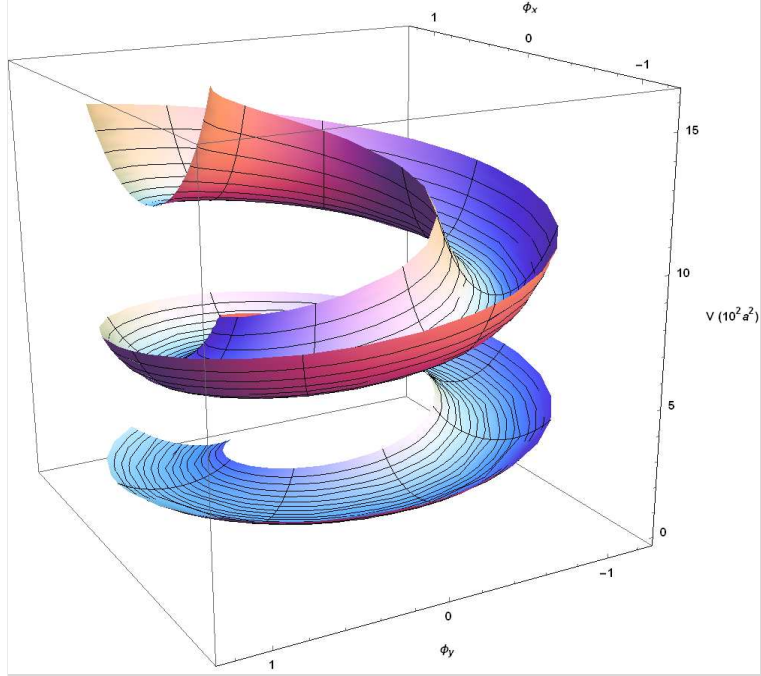}
\caption{The helicoid structure of potential (\ref{mini}) scaled by $10^2a^2$ . In the graph the parameters $b$ and $\Lambda$ are set to be $b=0.1$ and $\Lambda=1$ respectively.} \label{g}
\end{figure}

Without $\Lambda$, the superpotential comes back to (\ref{sup}) and the scalar potential is shown in Fig.~\ref{g} with $b=0.1$. During inflation $\langle r\rangle\simeq r_0$ for small $b$, the term $(\ln r)^2$ contribution to the potential, at the lowest order, is proportional to $b^2$. After canonical field nomalization, the inflaton potential takes the form
\begin{equation}
V(\te)=\frac{1}{2}m_\te^2(2b^2+\te^2),
\end{equation}
in which the higher order terms proportional to $b^{3+i}$ are ignored. Concern to the inflation observations, taking the tensor-to-scalar ratio \textbf{r} for example,
\begin{equation}
\textbf{r}=\frac{32\te_i^2}{(\te_i^2+2b^2)^2}\approx\frac{8}{N}(1+\frac{b^2}{2N})^{-2},
\end{equation}
where $\te_i$ is the phase when inflation starts, and $N\in(50, 60)$ is the e-folding number. So the correction from higher order term is insignificant for $b<1$.

\section{Monodromy in Supersymmetric Field Theory}

As discussed before, the phase inflation naturally leads to the phase monodromy (in mathematical sense) in the superpotential. The phase monodromy requires singularity, which means the superpotential proposed for the phase inflation should be an effective theory. It is preferred to show how such phase monodromy appears from a more ``fundamental'' theory at higher scale. In \cite{Li:2014vpa} the monodromy needed for phase inflation is realized based on the supersymmetric field theory, in which the monodromy relates to the soft breaking of a global $U(1)$ symmetry.

Historically the monodromy inflation as an attractive method to realize super-Planckian field excursion was first proposed, in a more physical sense, for axions arising from string compactifications \cite{Silverstein:2008sg}. In the
inflaton potential, the only factor that changes during axion circular rotation is from the DBI action of branes. In \cite{Choi:2014rja, Tye:2014tja, Kappl:2014lra}, the generalized axion alignment mechanisms are considered as a particular realization of
axion monodromy with the potential from non-perturbative effects. We will show that, such kind of axion monodromy can also be fulfilled by the superpotential phase monodromy \cite{Li:2014vpa}, even though it is not shown in the effective superpotential after integrating out the heavy fields. Furthermore, all the extra fields can be consistently stabilized.

The more ``fundamental'' field theory for the superpotential in (\ref{sup}) is
\begin{equation}
W_0=\sigma X\Psi(T-\delta)+Y(e^{-\alpha T}-\beta \Psi)+Z(\Psi\Phi-\lambda), \label{sup1}
\end{equation}
where the coupling constants for the second and third terms are taken to be 1
for simplicity, and a small hierarchy is assumed between the first term and the last two terms, i.e., $\sigma\ll1$.
The coupling $Ye^{-\alpha T}$ is assumed to be an effective description of certain non-perturbative effects. Similar forms can be obtained from D-brane instanton effect in type \MyRoman{2} string theory (for a review, see \cite{Blumenhagen:2009qh}), besides, the coefficient $\alpha\propto \frac{1}{f}\gg1$ in Planck unit, since $f\ll1$ is the decay constant and should be significantly lower than the Planck scale.
For the last two terms of $W_0$, there is a global $U(1)$ symmetry
\begin{equation}
\begin{split}
&\Psi\rightarrow\Psi e^{-iq\te} ~,~\\
&\Phi\rightarrow\Phi e^{iq\te} ~,~\\
&Y\rightarrow Y e^{iq\te} ~,~\\
&T\rightarrow T+i q\te/\alpha~,~ \label{u1}
\end{split}
\end{equation}
which is anomalous and explicitly broken by the first term. The phase monodromy of superpotential $W$ in (\ref{mono}) originates from the $U(1)$ rotation of $W_0$
\begin{equation}
\Psi\rightarrow\Psi e^{-i2\pi},  ~~ W_0\rightarrow W_0+i2\pi\sigma \frac{1}{\alpha}X\Psi.
\end{equation}

As shown in \cite{Li:2014vpa}, the supersymmetric field theory with superpotential $W_0$ admits the Minkowski vacuum at
\begin{equation}
\begin{split}
&\langle X\rangle=\langle Y\rangle=\langle Z\rangle=0, ~\langle T\rangle=\delta, \\
&\langle \Psi\rangle=\frac{1}{\beta}e^{-\alpha \delta},
\langle \Phi\rangle=\lambda \beta e^{\alpha \delta},
\end{split}
\end{equation}
with $\langle\Phi\rangle\gg\langle\Psi\rangle$ so that near the vacuum the masses of $Y$, $Z$, and $\Psi$ are much larger than $\Phi$, besides, the effective mass of $T$ is also large near the vacuum due to the large $\alpha$. The large $\alpha$ was an obstacle for natural inflation as it leads to the axion decays too small for inflation, while in this scenario the large $\alpha$ is helpful for phase inflation to stabilize the axion. Therefore, for the physical process at scale below the mass scale of three heavy fields, the only unfixed degree of freedoms are $X$ and $\Phi$, which can be described by an effective field theory with  three heavy fields integrated out. The coupling $\sigma X\Psi(T-\delta)$ is designed for inflation and hierarchically smaller than the extra terms in $W_0$.
Therefore, to describe inflation process, the heavy fields need to be integrated out.

To integrate out heavy fields, we need to consider the F-terms again.
The F-term flatnesses of fields $Y$ and $Z$ give
\begin{equation}
\begin{split}
&F_Y=e^{-\alpha T}-\beta \Psi+K_YW_0=0, \\
&F_Z=\Psi\Phi-\lambda+K_ZW_0=0.
\end{split} \label{YZ}
\end{equation}
Near the vacuum $Y=Z\approx0\ll M_P$,  the above supergravity corrections $K_{Y(Z)}W_0$ are ignorable, and then the F-term flatness conditions reduce to
these for global supersymmetry. This is benefited from the fact that although the inflation dynamics is subtle, the inflation energy density is close to the GUT scale, far below the Planck scale. Solving the F-term flatness equations in (\ref{YZ}), we obtain the effective superpotential $W$ in (\ref{sup}).

Based on above construction, it is clear that the phase monodromy in $W$ is from the $U(1)$ transformation of $W_0$, and the pole of superpotential (\ref{sup}) at $\Phi=0$ arises from the integration process. The heavy field $\Psi$ is integrated out based on the F-term flatness conditions when $\langle \Phi\rangle\gg\langle\Psi\rangle$, while if $\Phi\rightarrow0$, $\Psi$ becomes massless from $|F_Z|^2$ and it is illegal to integrate out a ``massless'' field.
For inflation the condition $\langle \Phi\rangle\gg\langle\Psi\rangle$ is satisfied so the theory with superpotential (\ref{sup}) is reliable.

As to the inflation term, a question appears that the global $U(1)$ is explicitly broken by the first term in (\ref{sup1}) at inflation scale, why the phase is light while the norm is much heavier? The supergravity correction to the scalar potential plays a crucial role at this stage. The coefficient $e^K$ appears in the scalar potential, and because of the $U(1)$ symmetry in the K\"ahler potential, the factor $e^K$ is invariant under $U(1)$ symmetry but increases exponentially for a large norm. Here, the K\"ahler potential of $T$ should be shift invariant, i.e., $K=K(T+\bar{T})$ instead of the minimal type. Otherwise, the exponential factor $e^K$ depends on phase as well and the phase rotation will be strongly fixed, like the norm component or $\text{Re}(T)$.

When integrating out the heavy fields, they should be replaced both in superpotential and K\"ahler potential by the solutions from vanishing F-term equations. So different from the superpotential, the K\"ahler potential obtained in this way is slightly different from the minimal case given in (\ref{kp}). There are extra terms like
\begin{equation}
\Psi\bar{\Psi}=\frac{\lambda^2}{r^2}, ~~ K(T+\bar{T})=K({\frac{1}{\alpha^2}(\ln r)^2}), \label{cor}
\end{equation}
where $|\Phi|=r$. Nevertheless, since $\lambda\ll1$ and $\alpha\gg1$, these terms are rather small and have little effect on phase inflation, as shown in the last Section. Furthermore, the quantum loop effects during integrating out heavy fields can introduce corrections to the K\"ahler potential as well. While because of the $U(1)$ symmetry built in the K\"ahler potential, these terms, together with (\ref{cor}), can only mildly affect the field stabilization, and the phase inflation is not sensitive to the corrections in K\"ahler potential. As to the superpotential, it is protected by the non-renormalized theorem and free from radiative corrections.

\section{Natural Inflation in Helical Phase Inflation}

In the superpotential $W_0$, the inflation term is perturbative coupling of complex field $T$ which shifts under the global $U(1)$, an interesting modification is to consider the inflation given by non-perturbative coupling of $T$. Such term gives a modified $U(1)$ phase monodromy in the superpotential.
And it leads to the natural inflation as a special type of helical phase inflation with phase-axion alignment,
which is similar to the axion-axion alignment mechanism proposed in \cite{Kim:2004rp} for natural inflation with super-Planckian axion decay constant. Specifically it can be shown that the phase monodromy realized in supersymmetric field theory has similar physical picture with the modified axion alignment mechanism provided in \cite{Choi:2014rja, Tye:2014tja}.

To realize natural inflation, the superpotential $W_0$ in (\ref{sup1}) just needs to be slightly modified
\begin{equation}
W_1=\sigma X\Psi(e^{-\alpha T}-\delta)+Y(e^{-\beta T}-\mu \Psi)+Z(\Psi\Phi-\lambda), \label{sup2}
\end{equation}
in which $1\ll\alpha\ll\beta$. Again there is a global $U(1)$ symmetry in the last two terms of $W_1$, and the fields transfer under $U(1)$ like in (\ref{u1}).
 The first term, which is hierarchically smaller, breaks the $U(1)$ symmetry explicitly, besides a shift symmetry of $T$ is needed in the K\"ahler potential. The monodromy of $W_1$ under a circular $U(1)$ rotation is
\begin{equation}
\Psi\rightarrow\Psi e^{-i2\pi},  ~~ W_1\rightarrow W_1+\sigma X\Psi e^{-\alpha T}(e^{-2\pi i\frac{\alpha}{\beta}}-1).
\end{equation}

The supersymmetric field theory given by (\ref{sup2}) admits the following supersymmetric Minkowski vacuum
\begin{equation}
\begin{split}
&\langle X\rangle=\langle Y\rangle=\langle Z\rangle=0, ~\langle T\rangle=-\frac{1}{\alpha}\ln \delta, \\
&\langle \Psi\rangle=\frac{1}{\mu}\delta^{\frac{\beta}{\alpha}},
~~\langle \Phi\rangle=\lambda \mu \delta^{-\frac{\beta}{\alpha}}.
\end{split}
\end{equation}
The parameters are set to satisfy the conditions
\begin{equation}
\delta^{\frac{\beta}{\alpha}}\sim \lambda\mu,~~\lambda\ll1, ~~\delta<1, \label{par}
\end{equation}
so that $\langle \Psi\rangle\ll\langle \Phi\rangle$ and $\langle T\rangle>0$.

Near vacuum the fields $Y, Z, \Psi$ and $T$ obtain large effective masses above inflation scale while $X, \Phi$ are much lighter. At the inflation scale the heavy fields should be integrated out. The F-term flatness conditions
 for fields $Y$ and $Z$ are
\begin{equation}
\begin{split}
&F_Y=e^{-\beta T}-\mu \Psi=0, \\
&F_Z=\Psi\Phi-\lambda=0,
\end{split} \label{YZ1}
\end{equation}
in which the supergravity corrections $K_{Y/Z}W_1$ are neglected as both $Y$ and $Z$ get close to zero during inflation. Integrating out $\Psi$ and $T$ from Eq.~(\ref{YZ1}), we obtain the effective superpotential $W^\prime$ from $W_1$
\begin{equation}
W^\prime=\sigma\lambda\frac{X}{\Phi}((\frac{\mu\lambda}{\Phi})^{\frac{\alpha}{\beta}}-\delta).
\end{equation}
Giving $\alpha\ll\beta$, the effective superpotential contains a term with fractional power. Inflation driven by complex potential with fractional power was considered in \cite{Harigaya:2014eta} to get sufficient large axion decay constant. Here the supersymmetric field monodromy naturally leads to the superpotential with fractional power, which arises from the small hierarchy of axion decay constants in two non-perturbative terms.

The helical phase inflation is described by the effective superpotential $W^\prime$, the role of phase-axion alignment is not clear from $W^\prime$ since it is hidden in the procedure of integrating out heavy fields.

The F-term flatness conditions (\ref{YZ1}) fix four degree of freedoms, for the extra degree of freedoms, they corresponds to the transformations
free from the constraints (\ref{YZ1})
\begin{equation}
\begin{split}
&\Psi\rightarrow\Psi e^{-u-iv} ~,~ \\
&\Phi\rightarrow\Phi e^{u+iv} ~,~ \\
&T\rightarrow T+u/\beta+iv/\beta~.
\end{split} \label{tr}
\end{equation}
The parameter $u$ corresponds to the norm variation of complex field $\Phi$, which is fixed by the supergravity correction on the scalar potential $e^{K(\Phi\pb)}$. The parameter $v$ relates to the $U(1)$ transformation, which leads to the phase monodromy from the first term of $W_1$. The scalar potential, including the inflation term, depends on the superpositions among phases of $\Psi$, $\Phi$ and the axion $\text{Im}(T)$, which are constrained as in (\ref{tr}). Among these fields, the phase of $\Phi$ has the lightest mass after canonical field
nomalization. Similar physical picture appears in the axion-axion alignment that the inflation is triggered by the axion superposition along the flat direction.

\begin{figure}
\centering
\includegraphics[width=100mm, height=90mm,angle=0]{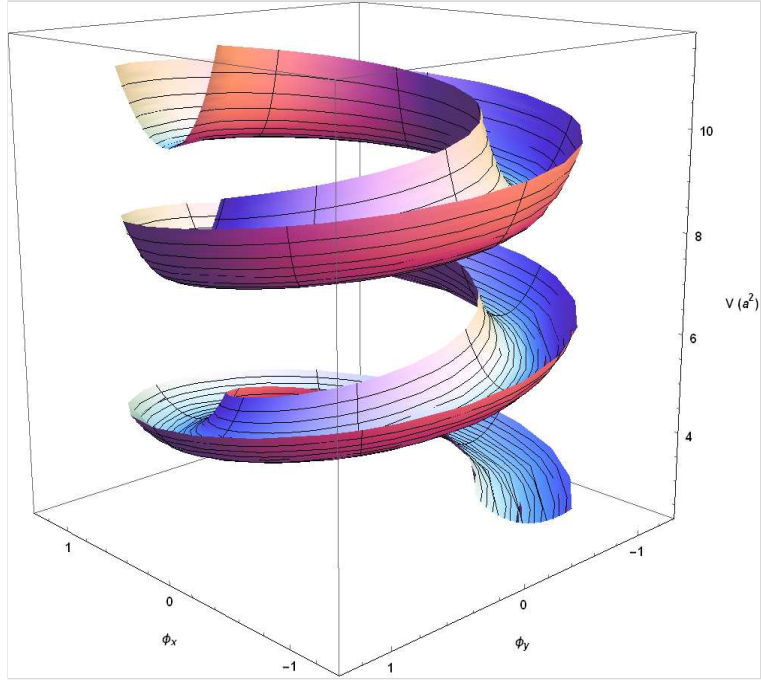}
\caption{The helicoid structure of potential (\ref{natp}) scaled by $a^2$.
The parameters with $c=0.96$ and $b=0.1$ are adopted in the graph. It is shown that the local valley locates around $r\approx1$. Note that the potential gets flatter at the top of the graph.} \label{g0}
\end{figure}

After integrating out the heavy fields, the K\"ahler potential is
\begin{equation}
K=\Phi\pb+\frac{\lambda^2}{\Phi\pb}+\cdots.
\end{equation}
As $\lambda\ll1$, the K\"ahler potential is dominated by $\Phi\pb$, as discussed before, the extra terms like $\frac{\lambda^2}{\Phi\pb}$ only give small corrections to the field stabilization. The helical phase inflation can be simply described by the supergravity
\begin{equation}
K=\Phi\pb+X\bar{X}+\cdots, ~~~~W=a\frac{X}{\Phi}(\Phi^{-b}-c),
\end{equation}
where $b=\frac{\alpha}{\beta}\ll1$ and $c\approx1$. The scalar potential given
 by the above K\"ahler potential and superpotential is
\begin{equation}
V=e^{r^2}\frac{a^2}{r^2}(r^{-2b}+c^2-2cr^{-b}\cos(b\te))~, \label{natp}
\end{equation}
in which we have used $\Phi \equiv re^{i\te}$. As usual the norm $r$ couples with the phase in the scalar potential, which makes it rather difficult to stabilize the norm while keep the phase light and then forbid
 the single field inflation. However, it is a bit different in helical phase inflation. The above scalar potential can be rewritten as follows
\begin{equation}
V=e^{r^2}\frac{a^2}{r^2}(r^{-b}-c)^2+e^{r^2}\frac{4a^2c}{r^{2+b}}(\sin\frac{b}{2}\te)^2~.
\end{equation}
So its vacuum locates at $\langle r\rangle=r_0=c^{-\frac{1}{b}}$ and $\te=0$. Besides, the coefficient of the phase term $e^{r^2}\frac{4c}{r^{2+b}}$ reaches its minimal value at $r_1=\sqrt{1+\frac{b}{2}}$. For $c\approx1$, we have the approximation $r_0\approx r_1\approx1$. The extra terms in $K$ give small corrections
 to $r_0$ and $r_1$, but the approximation is still valid. So with the parameters in (\ref{par}), the norm $r$ in the two terms of scalar potential $V$ can be stabilized at the close region $r\approx1$ separately. If the parameters are tuned so that $r_0=r_1$, then the norm of complex field is strictly stabilized at the vacuum value during inflation. Without such tuning a small difference between $r_0$ and $r_1$ is expected but the shift of $r$ during inflation is rather small and the inflation is still approximate to the single field inflation driven the phase term $\propto(\sin\frac{b}{2}\te)^2$. The helicoid structure of the potential (\ref{natp}) is shown in Fig.~\ref{g0}.

It is known that to realize aligned axion mechanism in supergravity \cite{Kallosh:2014vja, Higaki:2014pja}, it is very difficult to stabilize the moduli as they couples with the axions. In \cite{Li:2014lpa} the moduli stabilization is fulfilled with gauged anomalous $U(1)$ symmetries, since the $U(1)$ D-terms only depend on the norm $|\Phi|$ or $\text{Re}(T)$ and then directly separate the norms and phases of matter fields. In helical phase inflation, the modulus and matter fields except the phase are stabilized at higher scale or
by the supergravity scalar potential. Only the phase can be an inflaton
candidate because of the protection from the global $U(1)$ symmetry in K\"ahler potential and approximate $U(1)$ symmetry in superpotential.

\section{Helical phase inflation and the KYY model}
The $\eta$ problem in supergravity inflation can be solved both in the helical phase inflation and the shift symmetry in KYY model.
The physics in these two solutions are obviously different. For helical phase inflation, the solution employs the $U(1)$ symmetry in K\"ahler potential of the minimal supergravity, and the superpotential admits a phase  monodromy arising from the global $U(1)$ transformation
\begin{equation}
K=\Phi\pb+X\bar{X}+\cdots, ~~ W=a\frac{X}{\Phi}\ln\Phi. \label{ph}
\end{equation}
The inflation is driven by the phase of complex field $\Phi$ and several special virtues appear in the model. For the KYY model with shift symmetry \cite{Kawasaki:2000yn}, the K\"ahler potential is adjusted so that it admits a shift symmetry along the direction of $\text{Im}(T)$
\begin{equation}
K=\frac{1}{2}(T+\tb)^2+X\bar{X}+\cdots, ~~ W=aXT, \label{kyt}
\end{equation}
and the inflation is driven by the $\text{Im}(T)$. The shift symmetry is endowed with axions so this mechanism is attractive for axion inflation.

Here, we will show that, although the physical pictures are much different in helical phase inflation and KYY type model, just considering the lower order terms in the K\"ahler potential of redefined complex field, the helical phase inflation can reduce to the KYY model.

Because the phase of $\Phi$ rotates under the global $U(1)$ transformation, to connect the helical phase inflation with the KYY model, a natural guess is to take the following field redefinition $\Phi=e^T$, then the helical phase inflation (\ref{ph}) becomes
\begin{equation}
\begin{split}
&K=e^{T+\tb}+\cdots=1+T+\tb+\frac{1}{2}(T+\tb)^2+\cdots, \\
&W=aXTe^{-T}. \label{hs}
\end{split}
\end{equation}
The K\"ahler manifold of complex field $T$ is invariant under the holomorphic K\"ahler transformation
\begin{equation}
K(T,\tb)\rightarrow K(T,\tb)+F(T)+\bar{F}(\tb),
\end{equation}
in which $F(T)$ is a holomorphic function of $T$. To keep the whole Lagrangian also invariant under the K\"ahler transformation, the superpotential transforms under the K\"ahler transformation
\begin{equation}
W\rightarrow e^{-F(T)}W.
\end{equation}
For the supergravity model in (\ref{hs}), taking the K\"ahler transformation with $F(T)=-\frac{1}{2}-T$, the K\"ahler potential and superpotential become
\begin{equation}
K=\frac{1}{2}(T+\tb)^2+\cdots, ~~W=a\sqrt{e}XT,
\end{equation}
which is just the KYY model (\ref{kyt}) with higher order corrections in the K\"ahler potential. These higher order terms vanish after field stabilization and have no effect on inflation process.
The field relation $\Phi=e^T$ also gives a map between the simplest helical phase inflation and the KYY model, such as the inflaton: $\arg(\Phi) \rightarrow \text{Im}(T)$, and field stabilization $|\Phi|=1 \rightarrow\text{Re}(T)=0$.

Nevertheless, the helical phase inflation is not equivalent to the KYY model. There are higher order corrections to the K\"ahler potential in the map from helical phase inflation to KYY model, which have no effect on inflation after field stabilization but indicate different physics in two models. By dropping these terms certain information is lost so the map is irreversible. Specifically, the inverse function $T=\ln \Phi$ of the field redefinition $\Phi=e^T$ cannot reproduce the helical phase inflation from the KYY model. As there is no pole or singularity in the KYY model, it is unlikely to introduce pole and singularity at origin with phase monodromy through a well-defined field redefinition.
Actually, the singularity with phase monodromy in the superpotential indicates rich physics in the scale above inflation.

\section{Helical phase inflation in no-scale supergravity}
The no-scale supergravity is an attractive frame for GUT scale phenomenology \cite{Ellis:2013nka}, it is interesting to realize the helical phase inflation in no-scale supergravity.
Generally the K\"ahler manifold of the no-scale supergravity is equipped with $SU(N,1)/SU(N)\times U(1)$ symmetry. For the no-scale supergravity with exact $SU(1,1)/U(1)$ symmetry, without extra fields no inflation can be realized, so the case with $SU(2,1)/SU(2)\times U(1)$ symmetry is the simplest one that
admits inflation.

\begin{figure}
\centering
\includegraphics[width=100mm, height=90mm,angle=0]{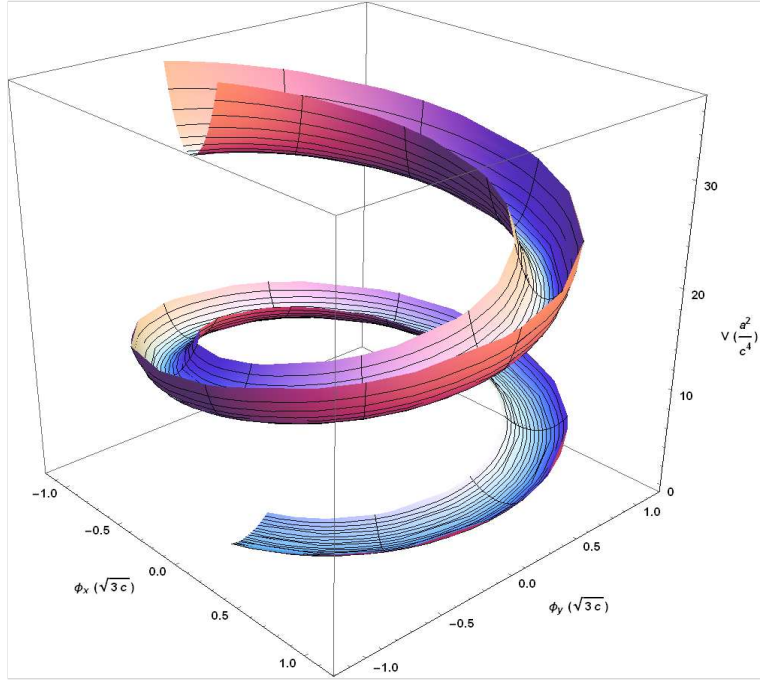}
\caption{The helicoid structure of potential (\ref{nos1}) scaled by $\frac{a^2}{c^4}$ . In the graph the parameter $\Lambda$ is tuned so that $r_0=r_1$. The complex field $\Phi$ has been rescaled by $\sqrt{3c}$, and the scale of field norm at local valley is determined by the parameter $c$ instead of the Planck mass.} \label{gg}
\end{figure}

The K\"ahler potential with $SU(2,1)/SU(2)\times U(1)$ symmetry is
\begin{equation}
K=-3\ln(T+\tb-\frac{\Phi\pb}{3}). \label{nos}
\end{equation}
In the symmetry of the K\"ahler manifold, there is a $U(1)$ subsector, the phase rotation of complex field $\Phi$, which can be employed for helical phase inflation. Besides, the modulus $T$ should be stabilized during inflation, which can be fulfilled by introducing extra terms on $T$. As a simple example of no-scale helical phase inflation, here we follow the simplification in \cite{Ellis:2013xoa} that the modulus $T$ has already been stabilized at $\langle T\rangle=c$.
Different from the minimal supergravity, the kinetic term given by the no-scale K\"ahler potential is non-canonical
\begin{equation}
L_K=K_{\Phi\pb}\partial_\mu\Phi\partial^\mu\pb=\frac{2c}{(2c-r^2/3)^2}(\partial_\mu r\partial^\mu r+r^2\partial_\mu\te
\partial^\mu\te),
\end{equation}
in which $\Phi \equiv re^{i\te}$ is used. The F-term scalar potential is
\begin{equation}
V=e^{-\frac{2}{3}K}|W_\Phi|^2=\frac{|W_\Phi|^2}{(T+\tb-\frac{\Phi\pb}{3})^2},
\end{equation}
where the superpotential $W$ is a holomorphic function of superfield $\Phi$ and $T=\langle T\rangle=c$. It requires a phase monodromy in superpotential $W$ for phase inflation, the simple choice is
\begin{equation}
W=\frac{a}{\Phi}\ln\frac{\Phi}{\Lambda}.
\end{equation}
The scalar potential given by this superpotential is
\begin{equation}
V=\frac{9a^2}{(6c-r^2)^2r^4}((\ln\frac{r}{e\Lambda})^2+\te^2)~. \label{nos1}
\end{equation}
As in the minimal supergravity, the norm and phase of complex field $\Phi$ is separated in the scalar potential. For the $r-$dependent coefficient factor $\frac{1}{(6c-r^2)^2r^4}$, its minimum locates at $r_0=\sqrt{3c}$ and another term $(\ln\frac{r}{e\Lambda})^2$ reaches its minimum at $r_1=e\Lambda$. Giving the parameter $\Lambda$ is tuned so that $r_0=r_1$,
in the radial direction the potential has a global minimum at $r=r_0$. Similar to the helical phase inflation in the minimal supergravity, the potential (\ref{nos1}) also shows helicoid structure, as presented in Fig.~\ref{gg}.

The physical mass of $r$ in the region near the vacuum is
\begin{equation}
m_r^2=\frac{(2c-r^2/3)^2}{4c}\frac{\partial^2 V}{\partial r^2}|_{r=\sqrt{3c}}=(4+\frac{1}{2\te^2})H^2,
\end{equation}
where $H$ is the Hubble constant during inflation.
So the norm $r$ is strongly stabilized at $r_0$. If $r_0$ and $r_1$ are not equal but close to each other, $r$ will slightly shift during inflation but its mass remains above the Hubble scale and the inflation is still approximately the single field inflation.

Instead of stabilizing $T$ independently with the inflation process, we can
 consider the helical phase inflation from the no-scale supergravity with dynamical $T$. There is a natural reason for such consideration. The phase monodromy requires matter fields transforming as rotations under $U(1)$, and also a modulus $T$ as a shift under $U(1)$. The K\"ahler potential of $T$ is shift invariant. Interestingly, for the no-scale supergravity with $SU(2,1)/SU(2)\times U(1)$ symmetry, as shown in (\ref{nos}), the K\"ahler potential $K$ is automatically endowed with the shift symmetry and global $U(1)$ symmetry
\begin{equation}
\begin{split}
&T\rightarrow T+iC, \\
&\Phi\rightarrow \Phi e^{i\te}.
\end{split}
\end{equation}
Therefore, the no-scale K\"ahler potential fits with the phase monodromy
in (\ref{sup1}) and (\ref{sup2}) initiatively.

The K\"ahler potentials of superfields $z\in{X, Y, Z}$ are of the minimal type $z\bar{z}$. While for $\Psi$, its K\"ahler potential can be the minimal type
$\Psi\bar{\Psi}$, or the no-scale type $K=-3\ln(T+\tb-(\Phi\pb+\Psi\bar{\Psi})/3)$, which extends the symmetry of K\"ahler manifold to $SU(3,1)/SU(3)\times U(1)$.
In this scenario, the process to integrate out heavy fields is the same as before. Besides, the potential of phase proportional to $|W_X|^2$ is insensitive to the formula of K\"ahler potential due to the $U(1)$ symmetry. The major difference appears in the field stabilization.
The scalar potential from phase monodromy in (\ref{sup1}) is
\begin{equation}
\begin{split}
V=e^K|W_X|^2=&\frac{1}{(\ln
r^2/\alpha+ c-r^2/3-\lambda^2r^2/3)^3}\frac{1}{r^2}((\ln r-\ln \Lambda)^2+\te^2), \\
\approx&\frac{1}{(c-r^2/3)^3}\frac{1}{r^2}((\ln r-\ln \Lambda)^2+\te^2),
\end{split}
\end{equation}
with $\alpha\gg1$ and $\lambda\ll1$.
The field norm $r$ is stabilized by minimizing the coefficient $\frac{1}{r^2(c-r^2/3)^3}$ and the parameter $c$ from $\langle T\rangle$ determines the scale of $\langle r\rangle=\frac{\sqrt{3c}}{2}$. This is different from the helical phase inflation in the minimal supergravity, which minimizes the norm based on the coefficient $e^{r^2}\frac{1}{r^2}$, and then the scale of $\langle r\rangle$ is close to $M_P$, the unique energy scale of supergravity corrections. Similarly, combining the no-scale K\"ahler manifold with phase monodromy in (\ref{sup2}), we can obtain the natural inflation.

\section{UV-sensitivity and Helical phase inflation}

For the slow-roll inflation, the Lyth bound \cite{Lyth:1996im} provides a relationship between the tensor-to-scalar ratio $\textbf{r}$ and the field excursion $\Delta\phi$. Roughly it requires
\begin{equation}
\frac{\Delta\phi}{M_P}\geqslant(\frac{\textbf{r}}{0.01})^{\frac{1}{2}}.
\end{equation}
To get large tensor-to-scalar ratio $\textbf{r}\geqslant0.01$, such as in chaotic inflation or natural inflation, the field excursion should be much larger
than the Planck mass. The super-Planckian field excursion makes the description based on the effective field theory questionable. In the Wilsonian sense, the low energy field theory is an effective theory with higher order corrections introduced by the physics at the cut-off scale, like quantum gravity, and these terms are irrelevant in the effective field theory since they are suppressed by the cut off energy scale. However, for the inflation process, the inflaton has
super-Planckian field excursion, which is much larger than the cut-off
 scale. Thus, the higher order terms can not be suppressed by the Planck mass and not irrelevant any more. And then they may significantly affect the inflation or even destroy the inflation process. For example, considering the following corrections
\begin{equation}
\Delta V=c_iV(\frac{\phi}{M_P})^{i}+\cdots, \label{hih}
\end{equation}
to the original inflaton potential $V$, as long as the $c_i$ are of
the order $10^{-i}$, in the initial stage of inflation $\phi\sim O(10)M_P$,
the higher order terms can be as large as the original potential $V$. So for large field inflation, it is sensitive to the physics at the cut-off scale and the predictions of inflation just based on effective field theory are questionable.

In consideration of the UV-sensitivity of large field inflation, a possible choice is to realize inflation in UV-completed theory, like string theory (for a review, see \cite{Baumann:2014nda}). To realize inflation in string theory, there are a lot of problems to solve besides inflation, such as the moduli stabilization, Minkowski/de Sitter vacua, and effects of $\alpha^\prime$- and string loop-corrections, etc. Alternatively, in the bottom-up approach, one may avoid the higher order corrections by introducing an extra shift symmetry in the theory. The shift symmetry is technically natural and safe under quantum loop corrections. However, the global symmetry can be broken by the quantum gravity effect. So it is still questionable whether the shift symmetry can safely evade
 the higher corrections like in (\ref{hih}).

The UV-completion problem is dodged in helical phase inflation. Since the super-Planckian field excursion is the phase of a complex field, and the phase component does not directly involve in the gravity interaction, there is no dangerous high-order corrections like in (\ref{hih}) for the phase potential. The inflaton evolves along the helical trajectory and does not relate to the physics in the region above the Planck scale. For the helical phase inflation in the minimal supergravity, the norm of field is stabilized at the marginal point of the Planck scale, where the supergravity correction on the scalar potential gets important based on which the norm of complex field can be strongly stabilized.
The extra corrections are likely to appear in the K\"ahler potential, however, they can only slightly affect the field stabilization while the phase inflation
is protected by the global $U(1)$ symmetry in the K\"ahler potential, 
in consequence the helical phase inflation is not sensitive to these corrections at all. For the helical phase inflation in no-scale supergravity, the norm of complex field is stabilized at the scale of the modulus $\langle T\rangle$ instead of the Planck scale, one can simply adjust the scale of $\langle T\rangle$ to keep the model away from super-Planckian region.

The helical phase inflation is free from the UV-sensitivity problem, and
 it is just a typical physical process at the GUT scale with special superpotential that admits phase monodromy.
So it provides an inflationary model that can be reliably studied just in supersymmetric field theory.

\section{Discussions and Conclusion}

In this work, we have studied the details of the helical phase inflation from several aspects. The helical phase inflation is realized in supergravity with global $U(1)$ symmetry. The $U(1)$ symmetry is built in the K\"ahler potential so
that the helical phase inflation can be realized by the ordinary K\"ahler potentials, such as the minimal or no-scale types. The helical phase inflation directly leads to the phase monodromy in the superpotential, which is singular and  an effective field theory arising from integrating out heavy fields.  The phase monodromy originates from the $U(1)$ rotation of the superpotential at higher scale. Generically, the superpotential can be separated into two parts $W=W_I+W_S$, where the $W_S$ admits the global $U(1)$ symmetry while $W_I$ breaks it
explicitly at scale much lower than $W_S$. Under $U(1)$ rotation the inflation term $W_I$ is slightly changed, which realizes the phase monodromy in the effective theory and introduces a flat potential along the direction of phase rotation. By breaking the global $U(1)$ symmetry in different ways, we may get
different kinds of inflation such as quadratic inflation, natural inflation,
 or the other types of inflation that are not presented in this work.

An amazing fact of helical phase inflation is that it deeply relates to several interesting points of inflation, and naturally combines them in a rather simple potential with helicoid structure. The features of helical phase inflation can be summarized as follows
\begin{itemize}
\item The global $U(1)$ symmetry is built in the K\"ahler potential, so the helical phase inflation provides a natural solution to $\eta$ problem.
\item The phase excursion requires phase monodromy in the superpotential. So the helical phase inflation provides, in the mathematical sense, a new type of monodromy in supersymmetric field theory.
\item The singularity in the superpotential, together with the supergravity
 scalar potential, provides a strong field stabilization which is consistent with phase inflation.
\item The super-Planckian field excursion is realized by the phase of a complex field instead of any other ``physical'' fields that directly couple
with gravity. So there is no polynomial higher order corrections for the phase and thus the inflation is not sensitive to the quantum gravity corrections.
\end{itemize}

To summarize, the helical phase inflation introduces a new type of inflation that can be effectively described by supersymmetric field theory at the GUT scale. Generically the super-Planckian field excursion makes the inflation predictions based on effective field theory questionable, since the higher order corrections from quantum gravity are likely to affect the inflation process significantly. One of the solution is to realize inflation in a UV-completed theory, like string theory, nevertheless, there are many difficult issues in string theory to
resolve before realizing inflation completely. The helical phase inflation is another simple solution to the UV-sensitivity problem.
It is based on the supersymmetric field theory and the physics is clear and much easier to control. Furthermore,
the helical phase inflation makes the unification of inflation theory with GUT more natural,
since both of them are happened at scale of $10^{16}$ GeV and can be effectively studied based on supersymmetric field theory.

Besides, we have shown that the helical phase inflation also relates to several interesting developments in inflation theory.
It can be easily modified for natural inflation, and realize the phase-axion alignment indirectly, which is similar to the axion-axion alignment mechanism for super-Planckian axion decay constant \cite{Kim:2004rp}. The phase-axion alignment is not shown in the final supergravity model which exhibits explicit phase monodromy only. However, the phase-axion alignment is hidden in the process when the heavy fields are integrating out.
For the $\eta$ problem in the supergravity inflation, there is another
well-known solution, the KYY model with shift symmetry.
We showed that through a field redefinition the helical phase inflation given by (\ref{kp}) and (\ref{sup}) reduces to the KYY model, where the higher order corrections in the K\"ahler potential have no effect on inflation process.
However, there is no inverse transformation from the KYY model to the helical phase inflation since no well-defined field redefinition can introduce the pole and phase singularity needed for phase monodromy. The helical phase inflation can be realized in no-scale supergravity. The no-scale K\"ahler potential automatically provides the symmetry needed by phase monodromy. In the no-scale supergravity, the norm of complex field is stabilized at the scale of the modulus.

Our inflation models are constructed within the supergravity theory with global $U(1)$ symmetry broken
 explicitly by the subleading order superpotential term. So it is just a typical GUT scale physics, and
indicates that a UV-completed framework seems to be not prerequisite to effectively describe such
inflation process.

\begin{acknowledgments}

The work of DVN was supported in part
by the DOE grant DE-FG03-95-ER-40917. The work of TL is supported in part by
    by the Natural Science
Foundation of China under grant numbers 10821504, 11075194, 11135003, 11275246, and 11475238, and by the National
Basic Research Program of China (973 Program) under grant number 2010CB833000.

\end{acknowledgments}

\end{document}